# Following the Long-Term Evolution of $sp^3$-type Defects in Tritiated Graphene using Raman Spectroscopy

G. Zeller [1,*], M. Schlösser [1] and H.H. Telle [2]

[1] Tritium Laboratory Karlsruhe (TLK), Institute for Astroparticle Physics (IAP), Karlsruhe Institute of Technology (KIT), Hermann-von-Helmholtz-Platz 1, 76344 Eggenstein-Leopoldshafen, Germany

[2] Departamento de Química Física Aplicada, Universidad Autónoma de Madrid, Campus Cantoblanco, 28049 Madrid, Spain

* Correspondence: genrich.zeller@kit.edu

## Abstract

We report on the evolution of tritium-induced $sp^3$-defects in monolayer graphene on a $Si/SiO_2$ substrate, by comparing large-area Raman maps of the same two samples, acquired just after fabrication and twice thereafter, about 9-12 months apart. Inbetween measurements the samples were kept under standard laboratory conditions. Using a conservative classification of $sp^3$-type spectra, based on the D/D' peak intensity ratio, we observed almost complete depletion of $sp^3$-type defects over the investigation period of about two years. This by far exceeds the ~5.5% annual reduction expected from tritium decay alone (≥ 3× larger). This change in the defect composition is accompanied by a recovery of the 2D-band of graphene and an overall decrease in defect-density, as determined via the D/G intensity ratio. Hydogenated graphene is reported to be reasonably stable over several months, when kept under vacuum, but suffers substantial hydrogen loss under laboratory air conditions. While the results shown here for tritiated graphene exhibit similarities with hydrogenated graphene, however, some distinct differences are observed.

## 1. Introduction

Graphene, a single atomic layer of carbon atoms in a honeycomb lattice, continues to be an intensively studied material, due to its exceptional electronic and mechanical properties.[1,2] One important property of graphene lies in the possibility of tuning its properties via the introduction of defects, or functional groups into/to the lattice.[3,4] One direct method of changing the lattice is through hydrogenation, whereby hydrogen atoms chemisorb and locally change $sp^2$-hybridized carbon into $sp^3$-hybridized configurations.[5–9] This disrupts the π-conjugated system, creates electronic gaps, and significantly alters electron transport and lattice-vibrational properties. Because these structural changes alter charge transport, phonon scattering, chemical reactivity, and barrier properties, hydrogenated and defect-engineered graphene appears across a broad range of applications from membranes and gas separation to catalysis, sensing, electronics, and protective coatings.[6,9–11]

Tritium, however, represents a qualitatively different case. As the radioactive isotope of hydrogen, it undergoes β-decay with a half-life of 12.3 years, emitting an energetic electron, an electron-antineutrino and leaving behind a $^3He$ daughter nucleus releasing a total decay energy of 18.6 keV.[12] The kinetic energy distribution of the electron is continuous with an average of 5.7 keV. Experiments with electron beams show that these energies are too low to create substantial damages to monolayer graphene.[13,14]

Like hydrogen and deuterium atoms, tritium atoms can chemisorb to graphene.[15–17] However, unlike the stable hydrogen/deuterium atoms, every tritium atom is an inherently unstable defect. After tritium decay, the $^3He$ daughter nucleus desorbs from the graphene surface as it cannot be chemically bound any longer.[18,19] Therefore, any tritiated graphene sample will lose about 5% of bound tritium within a year, just from the β-decay alone. In addition, the decay releases energy into the local lattice, which could potentially promote bond breaking, desorption of near-atoms, or create vacancies.

Thus, while hydrogen and deuterium functionalized systems that can remain relatively stable under appropriate storage conditions, tritiated graphene is expected to evolve with time, even under completely passive conditions. This combination of adsorption chemistry and nuclear instability makes tritium– graphene interactions a unique platform for studying defect dynamics; better understanding those is key for a number of applications and many technologies.

For example, graphene and graphene oxide membranes are intensively studied as selective filters for hydrogen isotope separation.[11] The usefulness of such membranes depends on their ability to maintain stable pore structures and functional groups.[13,20] Due to tritium exposure both the permeability and the isotope selectivity of the membrane could change with time, thus undermining long-term performance.

Similarly, graphene coatings are being investigated as ultra-thin barriers to hydrogen permeation in energy technology and tritium handling. In this case, decay-induced modification of the defect landscape could gradually open unwanted diffusion channels or reduce barrier effectiveness.[21,22]

A further possible application for tritiated graphene lies within the field of astroparticle physics. The PTOLEMY experiment[23,24] investigates the possibility to use atomic tritium bound on graphene, or graphene-like systems, as a target material to detect inverse β-decay from a potential cosmic neutrino background. In the KATRIN experiment[25] the possibility of tritium bound on graphene is explored as a potential low-activity, solid source of β-electrons from lattice-bound (atomic) tritium that could be used for the characterization of new detectors for future R&D.

The stability of the tritium–graphene system therefore represents itself as a fundamental open question in all applications of tritiated/tritium-exposed graphene. The question of stability is particularly relevant because it impacts the reproducibility of academic studies and the practical assessment of graphene's suitability in tritium-handling technologies.

The aim of this paper is to address precisely this point. We performed repeated Raman measurements on tritiated graphene sample, after storage of about one-to-two year under regular, non-specialist laboratory conditions. Using the so-called Eckmann model[26] for Raman defect analysis, we can not only quantify changes in the overall defect density but also demonstrate a clear shift in the dominant defect type over time.

## 2. Experimental section

The overall procedure of sample preparation, treatment, storage and analysis is shown in the schematic of Figure 1; experimental details are provided in the following segments.

### 2.1 Initial sample preparation

The samples used in this work are 1cm×1cm $SiO_2$/Si substrates (<100> Si, 525µm thick) with 90nm thermal $SiO_2$ on both sides and a monolayer of CVD-grown graphene on top.[27] According to the manufacturer, the graphene monolayer is transferred via 'polymer-assisted, semi-dry transfer'.[27] The general film quality is good, as characterized by the Raman intensity ratio $I_D / I_G < 0.1$ across the whole sample, with fluctuations of the order of 0.05.

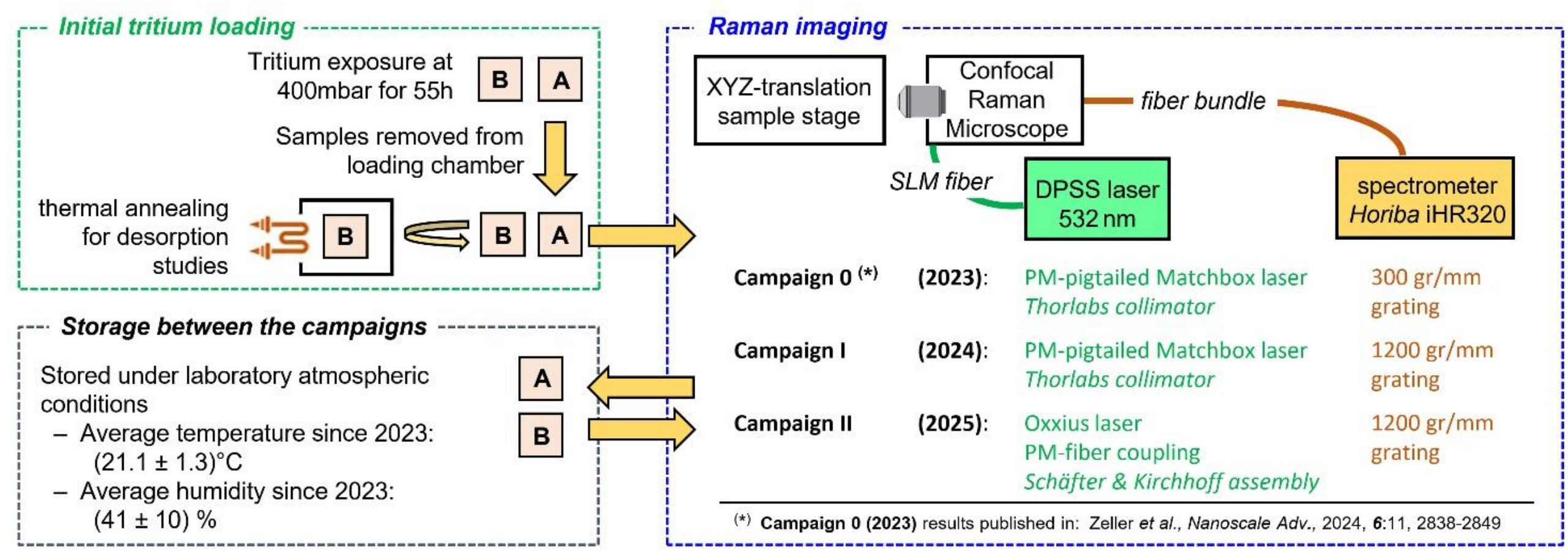

**Figure 1:** Schematic representation of the experimental procedure to prepare, store and analyse tritiated graphene (TG) samples; minor changes to the Raman imaging analysis between the measurement campaigns are indicated. Note that, the same samples were used for all measurement series (after fabrication and after being stored under laboratory atmospheric conditions for about 9-12 months in-between subsequent campaigns). For details see text.

The initial tritium loading[16] of the graphene samples was carried out in a custom-built tritium chamber with an internal volume of about 0.2L. Four monolayer graphene-on-$SiO_2$/Si samples (for distinction denoted as samples "A" to "D") were mounted on a stacked holder and exposed simultaneously. The central sample (sample B in this work) was contacted with four spring-loaded pins to allow for *in situ* sheet resistance measurements using the van der Pauw method[28] (Keithley DAQ6150 with 7709 matrix). The chamber was filled with a tritium gas mixture (97.2% $T_2$; remainder mainly HT/DT) at a pressure of ~400mbar, corresponding to a total activity of ~$7.6 \times 10^{12}$ Bq. The exposure time was ~55 h, after which the samples were removed for *ex situ* analysis. During the exposure to tritium the sheet resistance of the sample B increased from an initial sheet resistance of $R_s = (551 \pm 2)\ \Omega\ \square^{-1}$ to a plateau at $R_s \cong 129 \times 10^3\ \Omega\ \square^{-1}$, corresponding to a relative increase of ~250.

Two of these four samples (samples C and D) were destructively heated at 1400°C in a dedicated system, to determine the total amount of tritium that was adsorbed. It was found that the samples had acquired tritium equivalent to an activity of (19.0 ± 4.0) MBq. However, it is currently unknown, which fraction of tritium was adsorbed on the graphene itself and the substrate, respectively. A further sample (sample A) remained untreated after the exposure, and was stored as detailed in Section 2.2.

The sample B was subjected to additional studies during the initial work[16]. To investigate the reversibility of the changes introduced by tritium exposure, sample B was heated multiple times: first for 24.5h at 300°C, and subsequently for 22h at 500°C. After these treatments, it was stored as described in the next section.

## 2.2 Storage conditions

Samples A and B were stored under regulated laboratory conditions. Since 2023, the average TLK laboratory temperature was 21.1 (±1.3)°C, and the average relative humidity was 41 (±10) %.

Both before and after carrying out the actual Raman measurements (recording a full Raman map of any of the samples tended to take about one full day), the sample was always stored under regular laboratory air, for several months.

## 2.3 Raman Spectroscopy

The confocal microscope was equipped with a 10× objective lens (NA = 0.25), which results in a laser focal beam diameter on the graphene surface of 7.2 (±0.1) µm. The comparatively large beam spot is an inherent property our in-house built Raman microscope, constructed largely from *Thorlabs* components, for use with toxic and radioactive materials.[29] Since we work with macroscopic (1cm²) graphene samples, the large spot size is advantageous as it allows for faster, full-area sampling while still providing representative spectra with decent spatial resolution.

All Raman measurements were carried out using a 532nm excitation laser. The Campaign I data sets were recorded using a Matchbox® laser (*Integrated Optics*), at laser power of 120 mW (equating to a power density on the graphene surface of ≈ $2.9 \times 10^5$ W/cm²). The Campaign II data sets were recorded using a LaserBoxx LCX-532L unit (*Oxxius*), at 100mW output power (equating to a power density on the graphene surface of ~$2.4 \times 10^5$W/cm²). Even after prolonged exposure of several minutes at this power density, no changes or damage of the graphene sheet were observed, confirming that the measurement conditions are non-destructive.

It should be noted that, at the time of the initial analysis (denoted as "Campaign 0", see Figure 1), a high-resolution grating was not yet available.[16] These previous measurements with the low-resolution grating are therefore only briefly addressed in this present discussion. An upgrade to the high-resolution grating was performed in May 2024.

The Raman microscope is intensity-calibrated using the NIST SRM2242a standard, with calibration curves from 2023 and 2025 shown in Figure S1 of the *Supplementary Information.* Those curves demonstrate that the spectral sensitivity of the system has remained stable over several years of operation, despite hardware changes, such as a different new laser unit and a new spectrometer grating. This long-term reproducibility guarantees that, the data sets from the different campaigns can be directly, quantitatively compared, without introducing systematic bias associated with any instrumental changes.

Peak intensities and linewidths were obtained by fitting the respective Raman peaks with pseudo-Voigt functions during spectral analysis. Representative fit results are shown in Figure S2 of the *Supplementary Information.*

**Table 1**: Overview of the data sets discussed in this publication. The production and treatment history of the two studied samples (samples A and B) is summarised in the right-most column. TG = Tritiated Graphene; hTG = heated Tritiated Graphene.

| Dataset name | Sample | Measurement Date | History / Comments |
|---|---|---|---|
| TG-0 | Sample A | 23/07/2023 | - Tritium exposure for 55h at 400mbar<br>- Measured one week after sample preparation<br>- Only low-spectral resolution data available |
| TG-I | Sample A | 24/06/2024 | - Tritium exposure for 55h at 400mbar<br>- Stored in laboratory atmosphere for a total of 337 days |
| TG-II | Sample A | 17/07/2025 | - Tritium exposure for 55h at 400mbar<br>- Stored in laboratory atmosphere for a total of 725 days |
| hTG-I | Sample B | 03/07/2024 | - Tritium exposure for 55h at 400mbar in same batch as Sample A<br>- Heated afterwards for a total of 22h at 500°C<br>- Stored in laboratory atmosphere for a total of 337 days |
| hTG-II | Sample B | 10/06/2025 | - Stored in laboratory atmosphere for 688 days |

## 2.4 Measurement protocol and data sets

In this work, four data sets are analysed, as summarised in Table 1. These data sets were recorded from the two samples A and B, that were both simultaneously exposed to tritium, as described in Section 2.1. Accordingly, each sample contributes two data sets: one obtained in mid-2024 (Campaign I) and the other in mid-2025 (Campaign II).

A raster scan of sample A was performed on 24/06/2024 – this constitutes the *TG-I* (*TG* = Tritiated Graphene) data set. After 388 days, on 17/07/2025, another raster scan of sample A was carried out, using the same experimental parameters, resulting in the *TG-II* data set. The first raster scan of sample B was performed on 03/07/2024 – this constitutes the *hTG-I* (*hTG* = heated Tritiated Graphene) data set. After 342 days, on 10/06/2025, another raster scan of sample B was performed, using the same experimental parameters once again, resulting in the *hTG-II* data set.

A summary of the experimental line fit results is given in Table S1 in the *Supplementary information*, alongside representative Raman maps from which they were derived (Figure S3).

It should be noted that, the raster scans of samples A and B were performed across different-sized areas, with different step sizes. Because the samples had to be removed from the scan unit for storage, it was not possible to re-scan precisely the same areas in Campaign II as in Campaign I. However, all scans were consistently carried out within the central 3mm × 3mm region of each sample. Also, although we intended to remeasure both samples using identical parameters, this was prevented by the laser failure in late 2024. Thus, since the raster scans differed in sampled area and data set size, a uniform region of 40×40 data points (=1600 spectra) was extracted from each scan to allow for direct comparison of the statistical analysis results presented here.

# 3. Results and discussion

In pristine graphene, the Raman spectrum is dominated by the G-band (at ~1580cm$^{-1}$) and the 2D-band (at ~2670cm$^{-1}$). Both arise from phonon modes that do not require the presence of defects or disorder. A characteristic feature of high-quality graphene is that it exhibits an intensity ratio $I_G/I_{2D} < 1$.[30–32]

When defects are introduced, additional Raman bands appear. For hydrogenated / deuterated / tritiated graphene, the most relevant is the D-band (at ~1340cm$^{-1}$), which directly reflects defect activation.[9,33–35] The D′- band (at ~1620cm$^{-1}$) provides complementary information, as the intensity ratio $I_D/I_{D'}$ can be used to distinguish between sp³-type defects and vacancy-type defects.[26,36] In contrast to our previous publication[16], the D′-band is clearly resolved in the measurement results presented here, allowing the use of this ratio to assess the nature of the defects across the different data sets. Likewise, the intensity ratio $I_D/I_G$ is of key importance as it relates to the overall defect density of a graphene film.

It should be noted, however, that the D-band intensity is not strictly monotonic with defect density: starting in the low defect regime (in general addressed as "Stage 1") at very high levels of functionalisation, $I_D/I_G$ reaches a maximum before decreasing again ("Stage 2").[33,34]

The average Raman spectra of all four data sets and a pristine reference are shown in Figure 2.

Panel A shows the measurements recorded during Campaign I, while Panels B and C illustrate how the respective samples evolved between the two measurement Campaigns. In Panel B, which compares TG-I and TG-II, the 2D-peak intensity has recovered (from $I_D/I_G$=0.70 to $I_D/I_G$=1.28) to almost the level of pristine graphene. At the same time, the intensity of the D-peak decreased significantly (from $I_D/I_G$=3.51 to $I_D/I_G$=2.90), indicating a clear reduction of defect signatures during storage. While the average intensity of the D'-peak has not changed, the spread of the values has changed, which is discussed in detail later in this section. In contrast, Panel C shows that the hTG sample changed only minimally between the two measurement campaigns, with its spectral features remaining largely unchanged during storage.

At this point, for completeness, we ought to briefly address the Campaign 0 (TG-0) data set, which was recorded at low spectral resolution.[16] Due to the unresolved overlap of the G- and D'-peaks, neither the $I_{D'}/I_G$ nor $I_D/I_{D'}$ can be analysed and compared with Campaigns I and II. Instead, the D- and 2D-peaks may be used to make some qualitative comparisons and statements.[31,34] In Figure 3 the intensity ratio $I_D/I_{2D}$, and the widths $w_D$ and $w_{2D}$ are shown. The intensity ratio $I_D/I_{2D}$ for TG-0 is noticeably higher than for the other campaigns, suggesting that here the defect density was higher than in TG-I and TG-II. This is further supported by the analysis of the peak widths, which remain roughly unchanged between TG-I and TG-II, but are significantly larger for TG-0. This indicates that, at the time of the TG-0 measurements, the sample was in Stage 2 and had a higher defect density. Therefore, even though we cannot make quantitative statements about TG-0 and the defect evolution, it is clear that the defect density greatly diminished within the year between TG-0 and TG-I. However, the evolution of the $I_D/I_{2D}$ between Campaign I and II shows that even after the previous year of storage the change of the defect density is not yet concluded.

For the Campaign I and II data sets, we first demonstrate that the Eckmann model can be applied to distinguish between sp³-type and vacancy-type defects in our data sets. But it should be kept in mind that, as described earlier, this model is only valid in the low-defect regime of graphene. According to Cançado and co-workers, the width of the G-peak is often used to separate the Stage 1 regime from the Stage 2 regime.[34]

However, due to the overlap with the D′-peak, it is sometimes difficult to make a clear distinction. Therefore, based on the work Fournier and co-workers, we use the width of the D-peak instead, which is less affected by spectral overlap and provides a more robust indicator for the transition between the regimes.[36]

To visualise this notion, the $I_D/I_G$-ratio is plotted as a function of the D-peak width, $w_D$, for all four data sets; this is shown in Figure 4. Note that, all peak width data shown and discussed here have been derived by eliminating the instrumental broadening from the measured profiles.

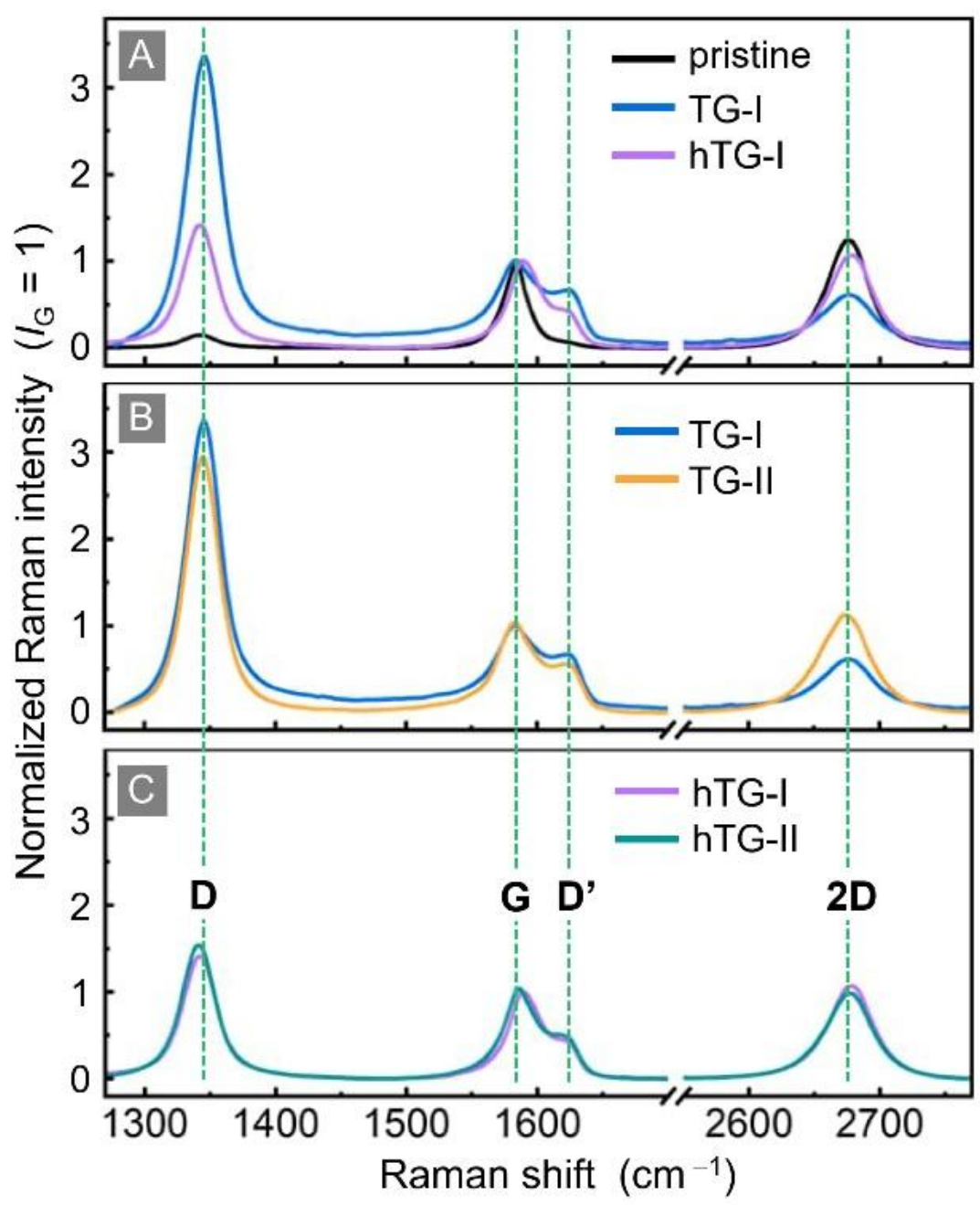

**Figure 2:** Comparison of the changes in Raman spectra of the different samples: pristine graphene (black trace); *TG-I*: $T_2$-exposed graphene (blue trace); *TG-II*: $T_2$-exposed graphene after one year of storage in laboratory atmosphere (orange trace); *hTG-I*: $T_2$-exposed graphene heated at 500°C for 24h (violet trace); *hTG-II*: heated $T_2$-exposed graphene after one year of storage in laboratory atmosphere (green trace). The (average) Raman spectra shown in panels (A) to (C) are normalized to the G-peak intensity. Key Raman spectral features are annotated.

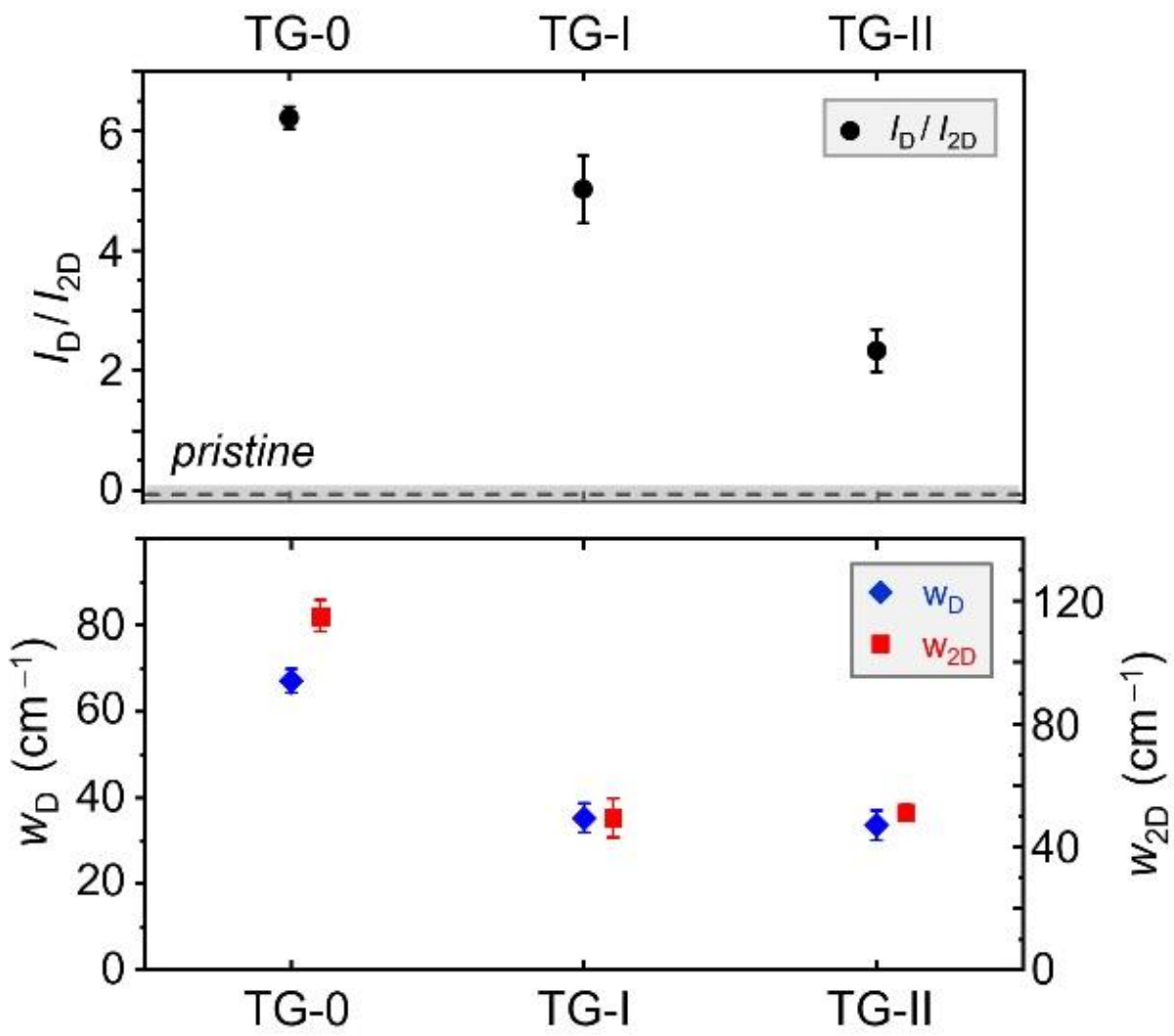

**Figure 3**: Comparison of the changes in Raman spectral features of the tritiated graphene (TG) samples throughout the different campaigns, including the data from Campaign 0 with lower spectral resolution. Top panel – intensity ratio $I_D/I_{2D}$ (black data points); the base value for the pristine graphene samples is marked by the dashed line (the error range is indicated by the grey shading). Bottom panel – width of the D-peak $w_D$ (blue data points) and width of the 2D-peak $w_{2D}$ (red data points).

Note that in data plots like the one used in the Figure 4, Raman spectra corresponding to Stage 1 accumulate in a band at lower widths. Given our spectral resolution, the Stage 1 band lies in the range $w_D$ = 25-35cm$^{-1}$; any data points with $w_D$ > 35cm$^{-1}$ correspond to Stage 2.

But note also that, this transition threshold value was determined empirically from the distribution of our data, but – nevertheless – is consistent with findings in the literature.[34,37] For easier visualisation, a kernel density estimation of the data points was performed, and contour lines enclosing 68%, 95%, and 99.7% (analogous to 1σ, 2σ, and 3σ intervals of a Gaussian distribution) of all 1600 data points were calculated. These contours, together with the mean values, are displayed in Figure 4 for all four data sets. Note that, the Eckmann model[26] can be applied to study the change of defect types between the respective data sets once the spectra are confirmed to fall into Stage 1. Data points outside this regime (Stage 2) are excluded from further analysis.

In addition to identifying the defect regimes, the contour plots reveal clear trends between the two campaigns. For TG-I and TG-II (panel A), both the average $I_D/I_G$-ratio and the width of the D-peak decrease. At the same time, the spread of data points narrows; this points to a gradual reduction and homogenisation of defect signatures during storage. For hTG-I and hTG-II (panel B), the average values remain more or less the unchanged, but the contour in the hTG-II plot reveal a “tail” towards larger $w_D$, suggesting that a fraction of spectra exhibits more disordered characteristics after storage. As visible in the Raman maps shown in the *Supplementary information* Figure S3 (C,D), the hTG-sample exhibits some hole-like structures and inhomogeneities that appear after the last heating cycle at 500 °C. We therefore assume that the “tail” is an effect of sampling a slightly different region of the hTG-sample.

To better visualise and investigate the change in defect types, we employed a variation of the plotting procedure established by Eckmann and co-workers.[26] In this adapted representation, the $I_D/I_G$-ratio is plotted as a function of the $I_{D'}/I_G$-ratio (see Figure 5).

Here, the slope of lines originating from (0,0) corresponds to the $I_D/I_{D'}$-ratio, which allows one to determine the different defect types. According to the Eckmann model, a ratio of $I_D/I_{D'} = 13$ corresponds to sp³-type defects, a ratio of $I_D/I_{D'} = 7$ to vacancy-type defects, and a ratio of $I_D/I_{D'} = 3.5$ to boundary-type defects. These respective reference lines are included in the display panels. Mixtures of defect types result in intermediate $I_D/I_{D'}$-ratios between the defect-specific reference values.

Similar to Figure 4, for clearer visualisation, a kernel density estimation of the data points was performed, and contour lines enclosing 68%, 95%, and 99.7% (analogous to 1σ, 2σ, and 3σ intervals of a Gaussian distribution) of the 1600 spectra were drawn. These contours, together with the mean value, are shown in the figure panels.

In the TG-I data set, the mean ratio is $I_D/I_{D'}$≈7, which corresponds to vacancy-type defects. At the same time, a significant fraction of data points lies closer to $I_D/I_{D'}$=13, indicating local hotspots of sp³-type defects. Comparing the TG-I with the hTG-I data sets reveals that, heating removes the sp³-type defects: the mean shifts to about $I_D/I_{D'}$≈4.5, and no data points exceed $I_D/I_{D'}$=7. This indicates that only vacancy- and boundary-type defects remain. A similar trend is observed when comparing the TG-I with the TG-II data sets. After one year of storage, the mean remains at $I_D/I_{D'}$~7, however the spread of data points becomes much smaller, and all data points lie below $I_D/I_{D'}$=12. This suggests that most of the $sp^3$-type defects present in TG-I have disappeared during storage.

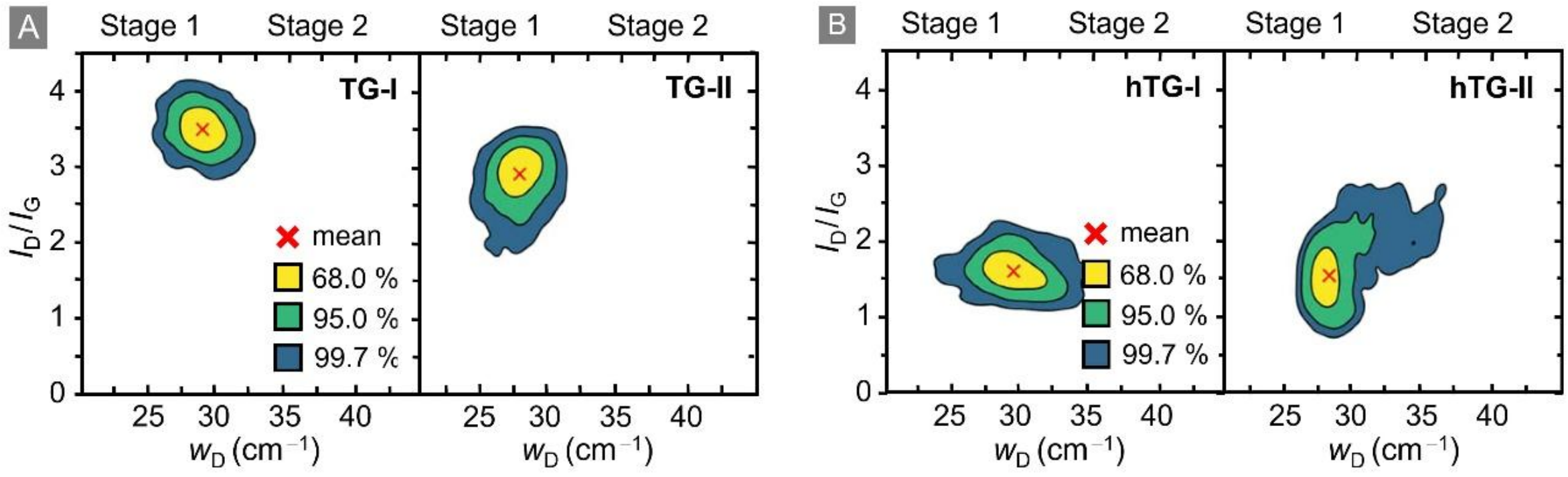

**Figure 4:** $I_D/I_G$ intensity ratio versus width $w_D$ of the D-peak. Based on Fournier et al.[37] this representation allows one to easily distinguish between Stage 1 and Stage 2 of the corresponding Raman data. The data points correspond to the surface areas of the respective samples shown in Figure S1 of the *Supplementary information*. The coloured contour regions in the figure correspond to 68 %, 95 %, and 99.7 % of all data points, respectively. The red cross indicates the mean value. (A) *TG-I*: $T_2$-exposed graphene, *TG-II*: $T_2$-exposed graphene after one year of storage in laboratory atmosphere; (B) *hTG-I*: $T_2$-exposed graphene heated at 500 °C for 24h, *hTG-II*: heated $T_2$-exposed graphene after one year of storage in laboratory atmosphere.

A quantitative description of the evolution of the defect types and numbers is difficult, since most of the data points fall into the range ~7 to ~13 for the $I_D/I_{D'}$-ratio; this means that – according to the Eckmann model – they indicate a mixture of defect types. At the time of writing, we were not aware of any established model to determine the individual contributions of the different defect types. However, knowledge of how to de-convolving the individual contributions would be crucial for the calculation of defect density, since the different defect types have different impact on the actual, measured $I_D/I_G$-ratio (which is a measure of the total number of defects).

In their model Lucchese[33] and Cançado[34] describe the evolution of the $I_D/I_G$-ratio as a function of the defect density, $L_D$, where $L_D$ is the mean distance between defects. However, this model is only valid for vacancy-type defects. In a recent publication Fournier and co-workers[37] established a model specifically for $sp^3$-defects. Note that, according to those models, Raman spectroscopy is on average almost two orders of magnitude more sensitive to vacancy-type defects than to $sp^3$-defects. To give a numerical example, in the so-called Lucchese/ Cançado model for vacancy-type defects the $I_D/I_G$-ratio maximum ($I_D/I_G \cong 3.6$) is reached for a defect density of about $L_D$~3nm. In contrast – according to the Fournier model – the same density of $sp^3$-type defects, the $I_D/I_G$-ratio would be substantially smaller, yielding about $I_D/I_G$~0.2. In this context, it should be noted that, the Lucchese/Cançado model has known normalisation problems and therefore cannot be used to infer reliable, quantitative statements about the defect density. Nevertheless, we opt to report values based on the Lucchese/Cançado model to allow direct comparison to prior studies, which used the same model formulation.

In Table 2 we collated values for the derived defect densities based on both models, for all four data sets. A summary of the calculation procedure can be found in *Supplementary information* S3. For example, for the TG-I data set, we obtain $L_D$=(0.58 ± 0.03) nm (applying the uncorrected Fournier model) and $L_D$=(4.69 ± 0.38) nm (applying the Lucchese/Cançado model); the latter value would be valid for only-vacancy-type defects, while the former would apply for only-$sp^3$-type defects. Based on the numbers reported in Table 2, specifically the average $I_D/I_G$-ratio, one can deduce that, within the year of storage between measurements, i.e., TG-I to the TG-II, the overall defect density has reduced by ~17.4 %.

In addition, this change is more pronounced when one analyses the change of defect types in detail. In Figure 6 overlayed histograms of the $I_D/I_G$-ratio of the TG-I and TG-II data sets are displayed. More than 60% of TG-I data points contain at least partial $sp^3$-type defects ($I_D/I_G$-ratio > 7), while just 0.8% of data points are classified as being associated with only-$sp^3$-type defects ($I_D/I_G$-ratio > 12).

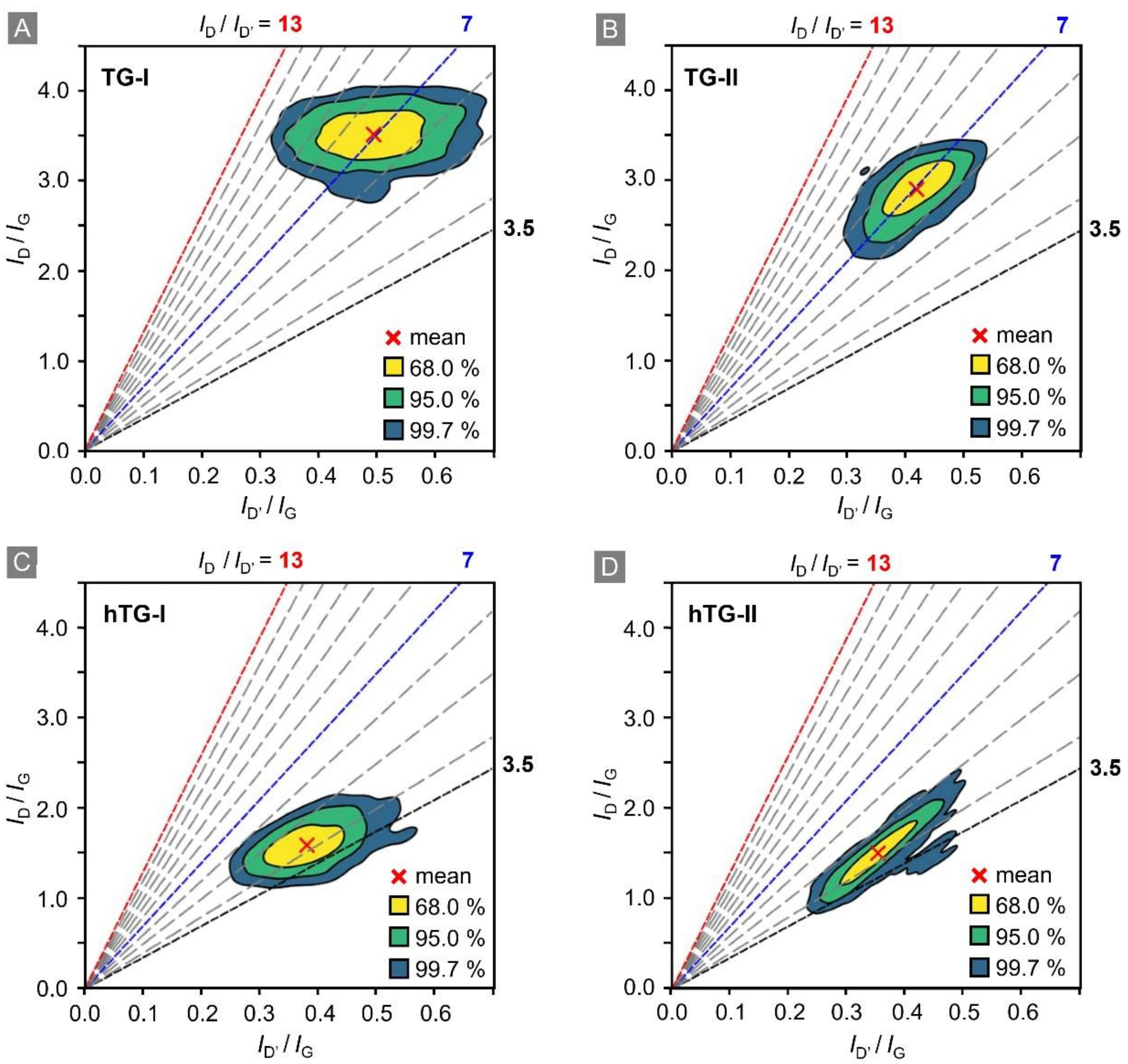

**Figure 5:** $I_D/I_G$ intensity ratio versus $I_{D'}/I_G$ intensity ratio; the slope-lines correspond to specific $I_D/I_{D'}$ intensity ratio. Based on the model by Eckmann *et al*[26] lines for the different types of defects are highlighted as following: boundary-type ($I_D/I_{D'} = 3.5$, black), vacancy-type ($I_D/I_{D'} = 7$, blue), and sp³-type ($I_D/I_{D'} = 13$, red). The data points correspond to the areas of the respective samples shown in Figure 2. The coloured contour regions in the figure correspond to 68 %, 95 %, and 99.7 % of all data points, respectively. The red crosses indicate the mean value. (A) *TG-I*: $T_2$-exposed graphene; (B) *TG-II*: $T_2$-exposed graphene after one year of storage in laboratory atmosphere; (C) *hTG-I*: $T_2$-exposed graphene heated at 500 °C for 24 h; (D) *hTG-II*: heated $T_2$-exposed graphene after one year of storage in laboratory atmosphere.

This large contribution of vacancy-type defects was already observed in the initial measurements (Campaign 0); therefore, we suspect that this might be a side effect of tritium exposure of graphene.[27] In contrast, in TG-II a much lower fraction of data points (about 22%) falls into the mixed category "vacancy+sp³" defects ($I_D/I_G$-ratio > 7); only-sp³-defects were not any longer observed. Therefore, Figure 6 visualises that, after two years a significant transformation of defect types has occurred, and sp³-type defects seem to have been largely depleted.

If radioactive decay were the only or dominant removal pathway, the expected 12-month loss would be ~5.5% (survival 0.945 for $T_{1/2}$=12.32y). Over the full investigation period of about two years this is much lower (≥ 3× total defect density, ≥ 10× depletion of sp³-defects) than the observed reduction, thus implying a combination of removal pathways. To correctly identify whether these pathways are tritium-specific, or are possible for all hydrogen isotopes, one needs a better understanding of the stability of hydrogenated (H,D,T) graphene in ambient conditions. However, because published studies examine almost invariably H-graphene on different substrates, in different environments, and for different sample preparations, one encounters vastly varying results on the stability of hydrogenated graphene.

**Table 2**. Defect densities for the different data sets based on the $I_D/I_G$- ratio and two different models. For the Lucchese/Cançado entries an uncorrected model is used – for details see main text.

| Data set | Average $I_D/I_G$-ratio | Defect density $L_D$ (nm) | |
|---|---|---|---|
| | | Lucchese/Cançado[33,34] | Fournier[37] |
| TG-I | 3.51 ± 0.26 | 4.69 ± 0.38 | 0.58 ± 0.03 |
| TG-II | 2.90 ± 0.28 | 5.61 ± 0.45 | 0.66 ± 0.04 |
| hTG-I | 1.60 ± 0.24 | 8.48 ± 0.79 | 0.95 ± 0.08 |
| hTG-II | 1.53 ± 0.29 | 8.71 ± 1.02 | 0.98 ± 0.11 |

Using XPS measurements, Apponi and co-workers[38] recently reported that, hydrogenated graphene is stable under vacuum conditions. In contrast, they report that, significant reversible oxidation occurs in laboratory air, with a saturation timescale of the order of just a few hours. Here it should be noted that using Raman spectroscopy it is difficult to reliably distinguish between hydrogenation and oxidation, since both give rise to $sp^3$-type defects.[26,36,39] Therefore, the oxidation reported by Apponi and co-workers cannot explain the observed depletion of $sp^3$-type defects in our tritiated graphene samples.

Kula and co-woekers[40] have also studied the stability of hydrogenated graphene, using resistance, Raman, and FTIR measurements in their analysis. They find that – in addition to some minor oxidation – hydrogenation of graphene is almost fully reversible by exposure of the sample to humid oxygen, or humid air. Using humid oxygen, repeatedly and reliably they could de-hydrogenated their sample to almost pristine conditions, with response times of ~10min. Therefore, although Kula reports that, hydrogenated graphene is unstable in ambient air, they report much shorter timescales than we observe for tritiated graphene. Alone the fact that, we can still measure $sp^3$-defects in TG-I, when the sample is already one year old (see Figures 1 and 3), suggests much longer $sp^3$-depletion time-scales.

In summary, we observe a much stronger reduction of defects than the 12-month loss of ~5.5%, as expected for radioactive $\beta$-decay of tritium. In fact, we find an almost full depletion of $sp^3$-type defects two years after the initial sample preparation. Due to the highly variable experimental conditions in the existing literature, we cannot definitively conclude if this is a general instability of hydrogen isotopes bound on graphene, or is actually influenced by tritium-specific pathways. To arrive at an unequivocal answer to this, one would need further, dedicated studies with both hydrogen and tritium under the same experimental conditions.

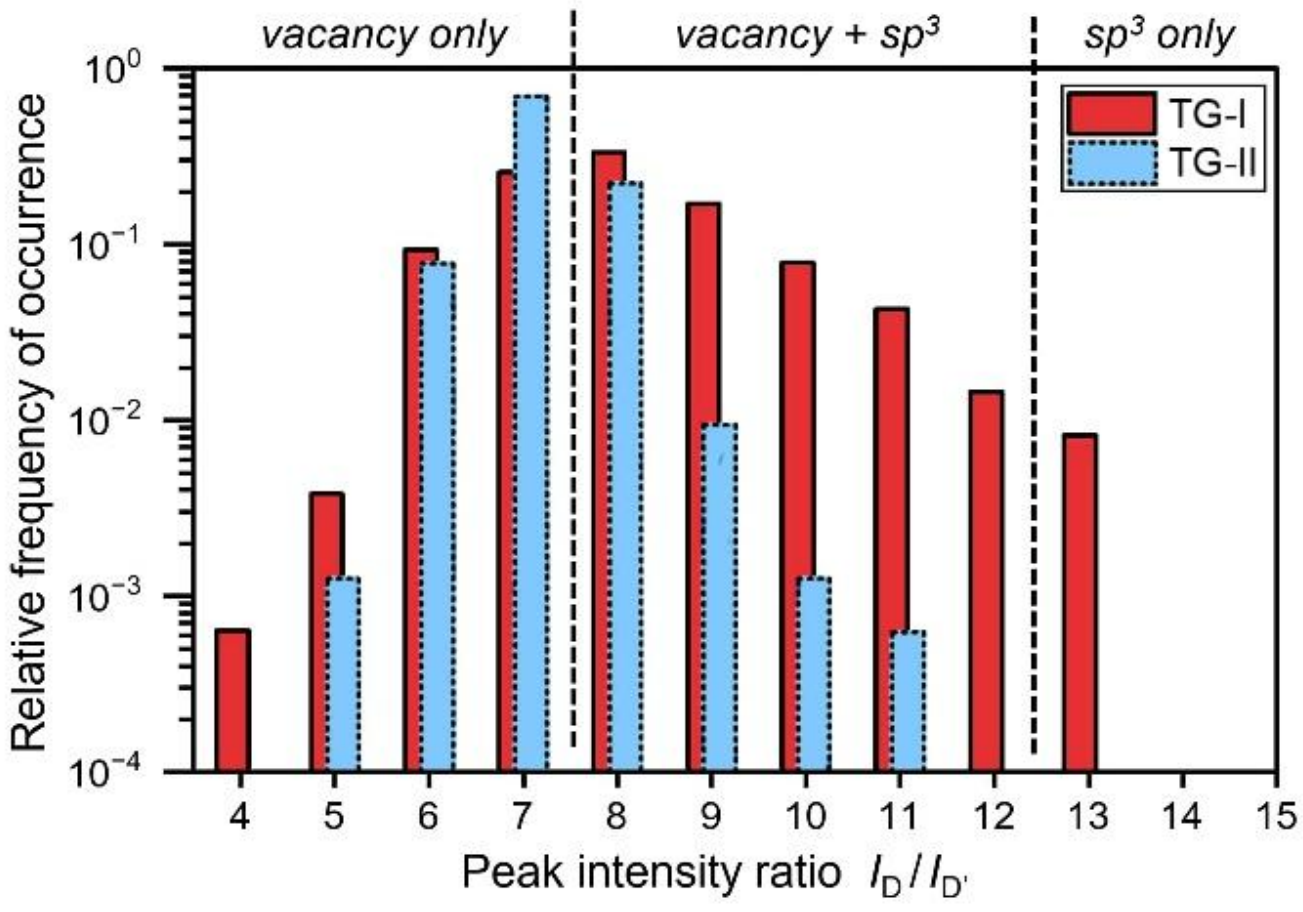

**Figure 6**: Histograms of the $I_D/I_G$-ratio of the TG-I and TG-II data sets, hinting at the defect type distribution. Defect-type regions, according to the Eckmann model, are annotated. For TG-I, more than 60% of all data points contain at least some $sp^3$-type defects, whereas for TG-II only <1% of data points contain $sp^3$-type defects.

# 4. Conclusions

Using Raman microscopy, we measured and re-remeasured the same tritiated-graphene samples, over the period of two years (with intermediate storage in laboratory air), to track – using the Eckmann model approach[26] – changes in defect chemistry between two data sets, TG-1 and TG-II.

The analysis shows a clear shift away from $sp^3$-type defects. If β-decay of bound tritium atoms were the only removal pathway, the expected 12-month loss would be around ~5.5% (corresponding to a survival rate of ~0.945). Instead, the reduction of the $sp^3$-classified data points is significantly larger, of the order of ≥10× higher than for the decay-only expectation. In contrast, the heated, de-tritiated control sample remains stable in both defect density and defect type over the same time interval.

Comparative studies by other research groups, albeit only for hydrogenated graphene, indicate that the stability strongly depends on the nature of the substrate and ambient-atmosphere chemistry. The general consensus is that, under vacuum conditions hydrogenated graphene seems to be rather stable, over the time scale of months. On the other hand, results for in-laboratory-air experiments are largely inconclusive, if not contradictory. Some studies report that, under air reversible oxidation of graphene samples is encountered (without significant de-hydrogenation), while others observe that, graphene in humid $O_2$/air does rapidly de-hydrogenate, on a time scale of minutes-to-hour. Because Raman spectroscopy is unable to reliably distinguish hydrogenation from oxidation (both processes generate $sp^3$-like signatures), oxidation reported for ambient air conditions does not, by itself, explain the sustained, long-timescale depletion we observe in our tritiated-graphene samples. Notably, measurable $sp^3$-signals persist in T-graphene (TG-I), even after about ~1 year storage under ambient air conditions. This seems to imply chemical kinetic processes that are slower than those reported for humid-gas de-hydrogenation of H-graphene.[40]

To attempt to resolve this conundrum, we are planning side-by-side H- and T-graphene experiments under identical, explicitly controlled conditions (substrate; vacuum vs. dry $O_2$ vs. humid $O_2$/air), including *in situ*, time-resolved tracking of sample properties, on time scales from just minutes up to several months. For this, a new tritium loading cell has been acquired that will allow for simultaneous *in-situ* sheet resistance measurements and *in-situ* Raman microscopy mapping, in real time. In parallel, we are exploring options of adding additional analysis methods, like XPS, which could enhance our understanding of the defect evolution – specifically distinguish between hydrogenation/tritiation and oxidation.

## Conflicts of interest

The authors declare no conflicts of interest.

## Data availability

The important data supporting the findings of this study are included within the article. Additional detailed data supporting this article have been included as part of the *Supplementary information* (SI).

## Acknowledgements

We like to thank S. Niemes, N. Tuchscherer and P. Wiesen for their contributions to the experimental setup and sample preparation in Campaign 0. In addition, we acknowledge the assistance of D. Díaz Barrero in the upgrade of the confocal Raman microscope. We also like to thank B. Bornschein for facilitating this work at TLK. Financial support for this CRM/graphene project has been provided by KCETA, and we acknowledge the seed funding by the KIT *Future Fields* programme through the “ELECTRON” project.

**Supplementary material**

# Following the Long-Term Evolution of $sp^3$-type Defects in Tritiated Graphene using Raman Spectroscopy

G. Zeller [1,*], M. Schlösser [1] and H.H. Telle [2]

[1] Tritium Laboratory Karlsruhe (TLK), Institute for Astroparticle Physics (IAP), Karlsruhe Institute of Technology (KIT), Hermann-von-Helmholtz-Platz 1, 76344 Eggenstein-Leopoldshafen, Germany

[2] Departamento de Química Física Aplicada, Universidad Autónoma de Madrid, Campus Cantoblanco, 28049 Madrid, Spain

* Correspondence: genrich.zeller@kit.edu

## S1 – CRM Raman spectral sensitivity calibration based on the NIST SRM 2242a standard

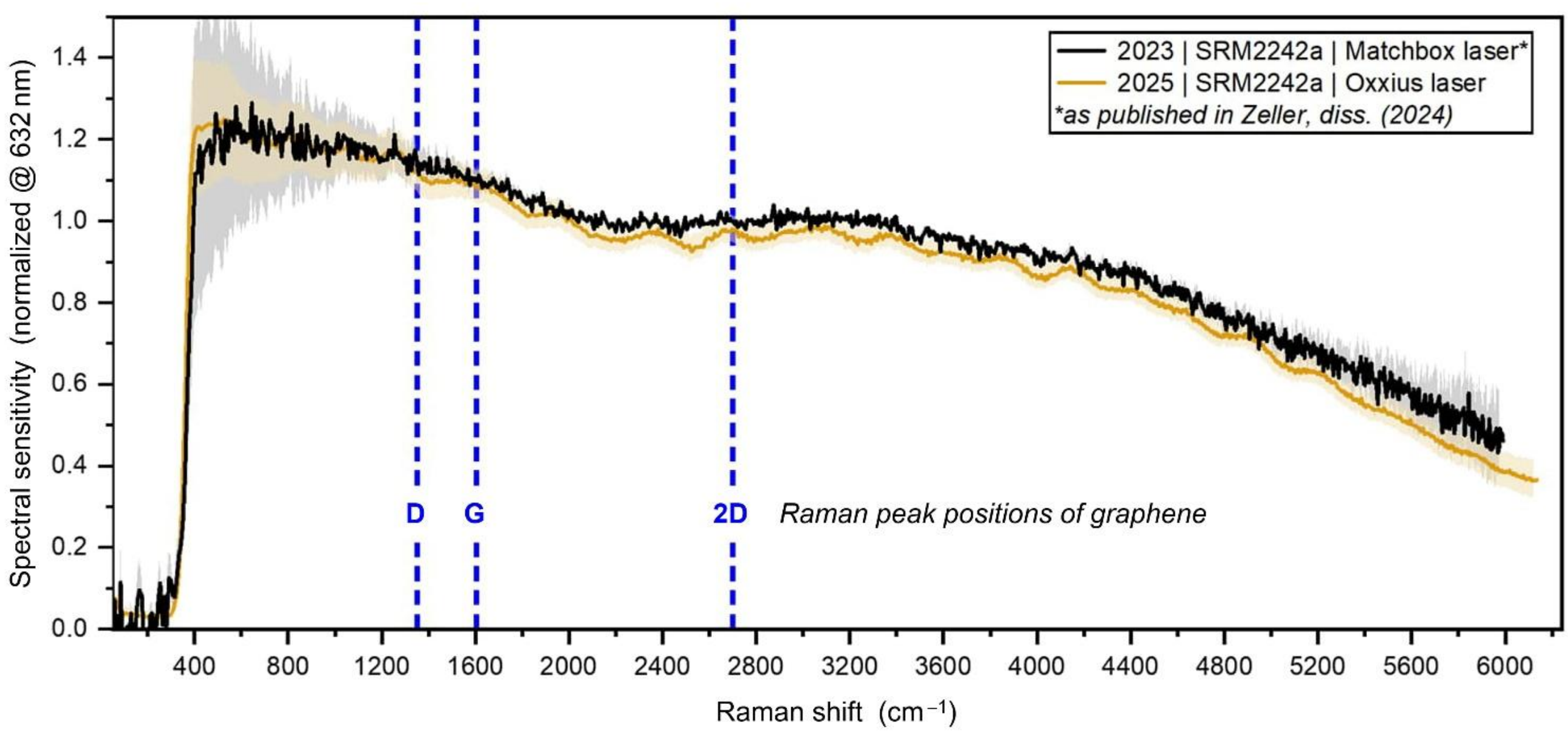

**Figure S1:** Long-term stability and comparability of the Raman system, based on response measurements using the NIST SRM 2242a standard.[41] The sensitivity curves for different years and hardware configurations (including new gratings and a replaced laser) agree within 1σ uncertainties, demonstrating year-over-year cross-comparability. The curves are normalized to =1 at the wavelength of the fluorescence emission maximum of the standard.

## S2 – Representative Raman spectra for all samples, with Voigt fits to Raman features

For the fit procedure, pseudo-Voigt profile functions were utilised.[42,43] In an often-used mathematical description they are given as the linear combination

$$V\big(x;\mathrm{w}_{(L)},\mathrm{w}_{(G)}\big) = \eta(r)\cdot L_V\big(x;\mathrm{w}_{(V)}\big) + \big(1-\eta(r)\big)\cdot G_V\big(x;\mathrm{w}_{(V)}\big)\ , \tag{1}$$

Its two common parameters are

- the width parameter $\mathrm{FWHM}_{(V)} \equiv \mathrm{w}_{(V)} \approx 0.5346\cdot\mathrm{w}_{(L)} + \sqrt{\mathrm{w}_{(G)} + 0.2166\cdot\mathrm{w}_{(L)}^2}$ , and (2)
- the mixing parameter $\eta(r) = 1.36603\cdot r - 0.47719\cdot r^2 + 0.11116\cdot r^3$, with $r \equiv \frac{\mathrm{w}_{(L)}}{\mathrm{w}_{(V)}}$ . (3)

Note that, in our fit implementation, the Lorentz ($L$) and Gaussian ($G$) component functions are taken in unit-height form and share the common width parameter, $\mathrm{w}_{(V)}$; thus, these line-shape functions are defined as

$$L\big(x;\mathrm{w}_{(V)}\big) = \frac{\mathrm{w}_{(V)}^2}{4\cdot(x-x_c)^2 + \mathrm{w}_{(V)}^2} \quad \text{and} \quad G\big(x;\mathrm{w}_{(V)}\big) = \exp\left(-\frac{(x-x_c)^2}{2\sigma_{(G)}^2}\right). \tag{4}$$

Here the Gaussian component of $V\big(x;\mathrm{w}_{(L)},\mathrm{w}_{(G)}\big)$ is associated with the resolution of the spectrometer, whose $\mathrm{FWHM}_{(G)}$ is evaluated from the analysis of the atomic line emission from a Ne-lamp (used for wavelength calibration of the spectrometer). For the spectrometer configuration used in campaign 0 this was $\mathrm{FWHM}_{(G)}(0) = 32 \pm 4\ \mathrm{cm}^{-1}$, while for campaigns I and II it was $\mathrm{FWHM}_{(G)}(\mathrm{I,II}) = 9.6 \pm 0.7\ \mathrm{cm}^{-1}$.

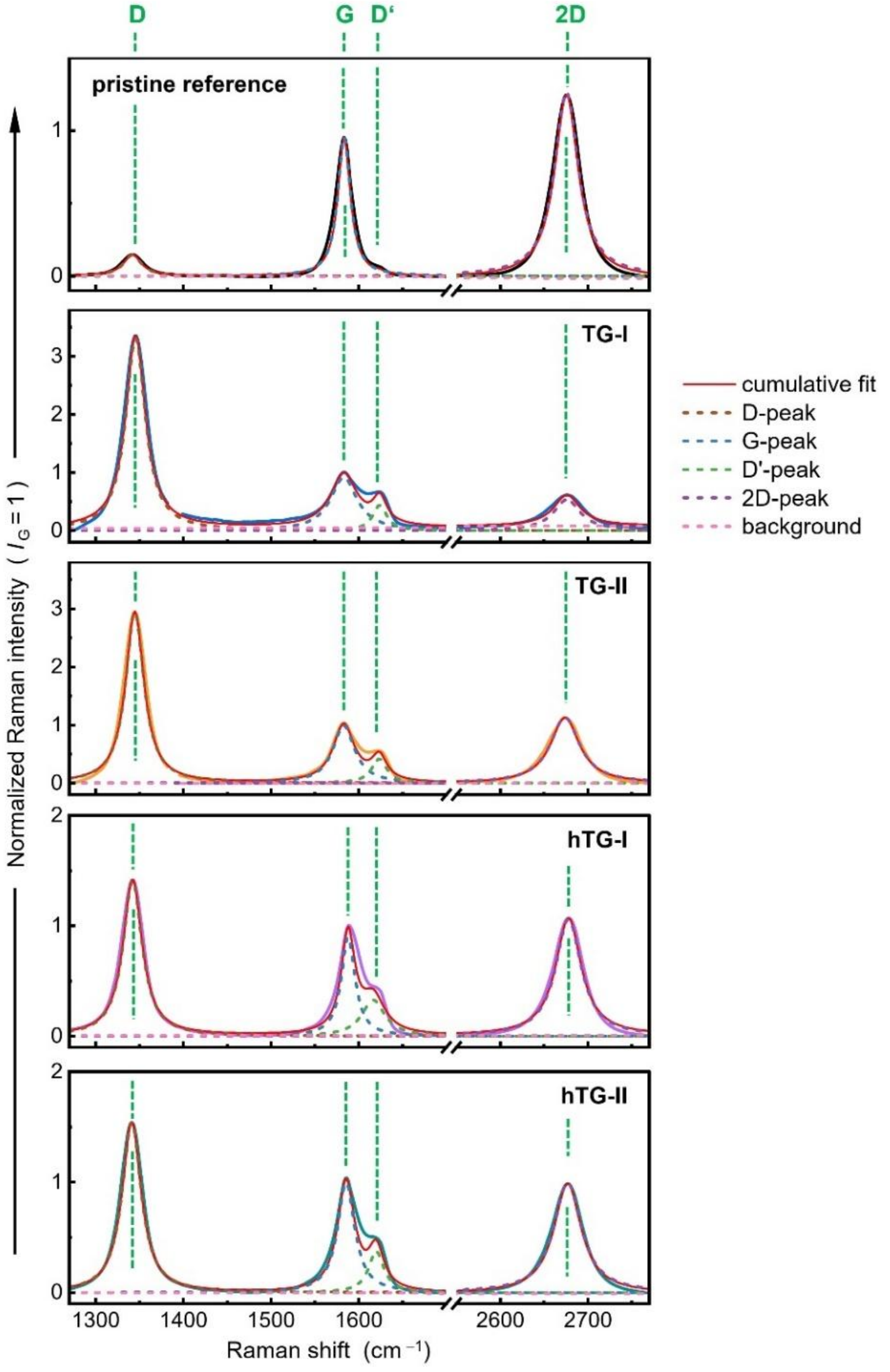

**Figure S2**: Representative Raman spectra of all sample data sets (pristine, TG-I, TG-II, hTG-I and hTG-II), shown with individual Voigt peak-fits and cumulative fit results. TG = tritiated graphene; hTG = heated tritiated graphene.

## S3 – Information extracted from the Raman spectroscopic maps of the samples

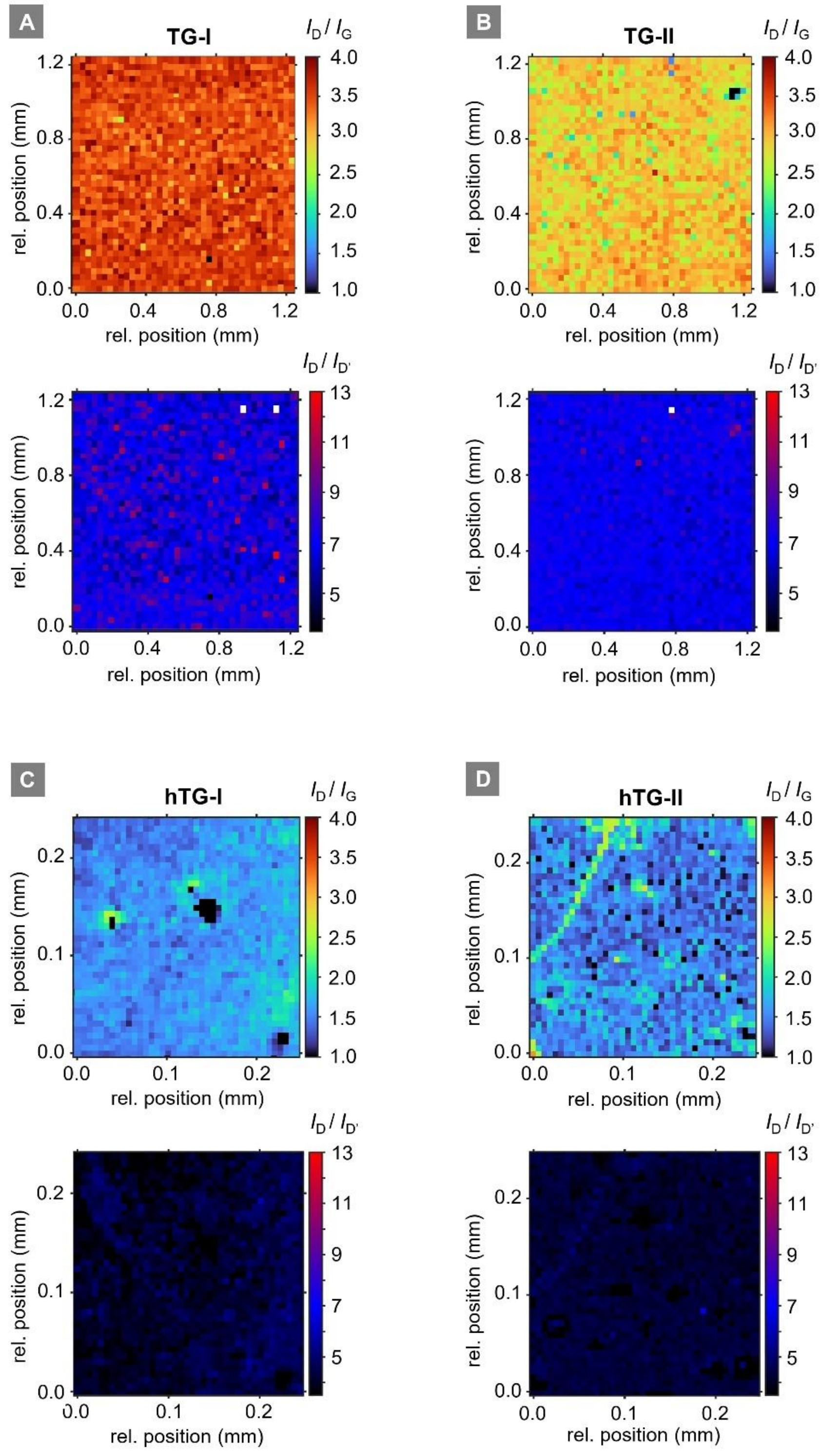

**Figure S3**: Raman spectroscopic maps of all graphene samples. Note that the one-year-later maps were not recorded for exactly equal areas of the samples. Top rows: $I_D/I_G$ intensity ratios; bottom rows: $I_D/I_{D'}$ intensity ratios.

(A) 1.2 × 1.2 mm² area scan with step size of ΔS = 32 µm of $T_2$-exposed graphene.

(B) 1.2 × 1.2 mm² area scan with step size of ΔS = 32 µm of $T_2$-exposed graphene, after one year of storage in laboratory atmosphere.

(C) 0.25 × 0.25 mm² area scan with step size of ΔS = 7 µm of $T_2$-exposed graphene, heated at 500 °C for 24 h.

(D) 0.25 × 0.25 mm² area scan with step size of ΔS = 7 µm of heated $T_2$-exposed graphene, after one year of storage in laboratory atmosphere.

### *S3.1 Numerical values for all Raman map points, and their graphical representation*

From the line ratio maps for the D-, D'- and G-peaks displayed in Figure S3 one can extract the actual numerical values, and plot these in the form $I_D / I_G$ versus $I_D / I_{D'}$. This plot is shown in Figure S4. Note that Figure 4 in the main text is a simplified representation of these data, in the form of a contour plot, lumping data points together around the mean (as 1σ-, 2σ- and 3σ contours regions).

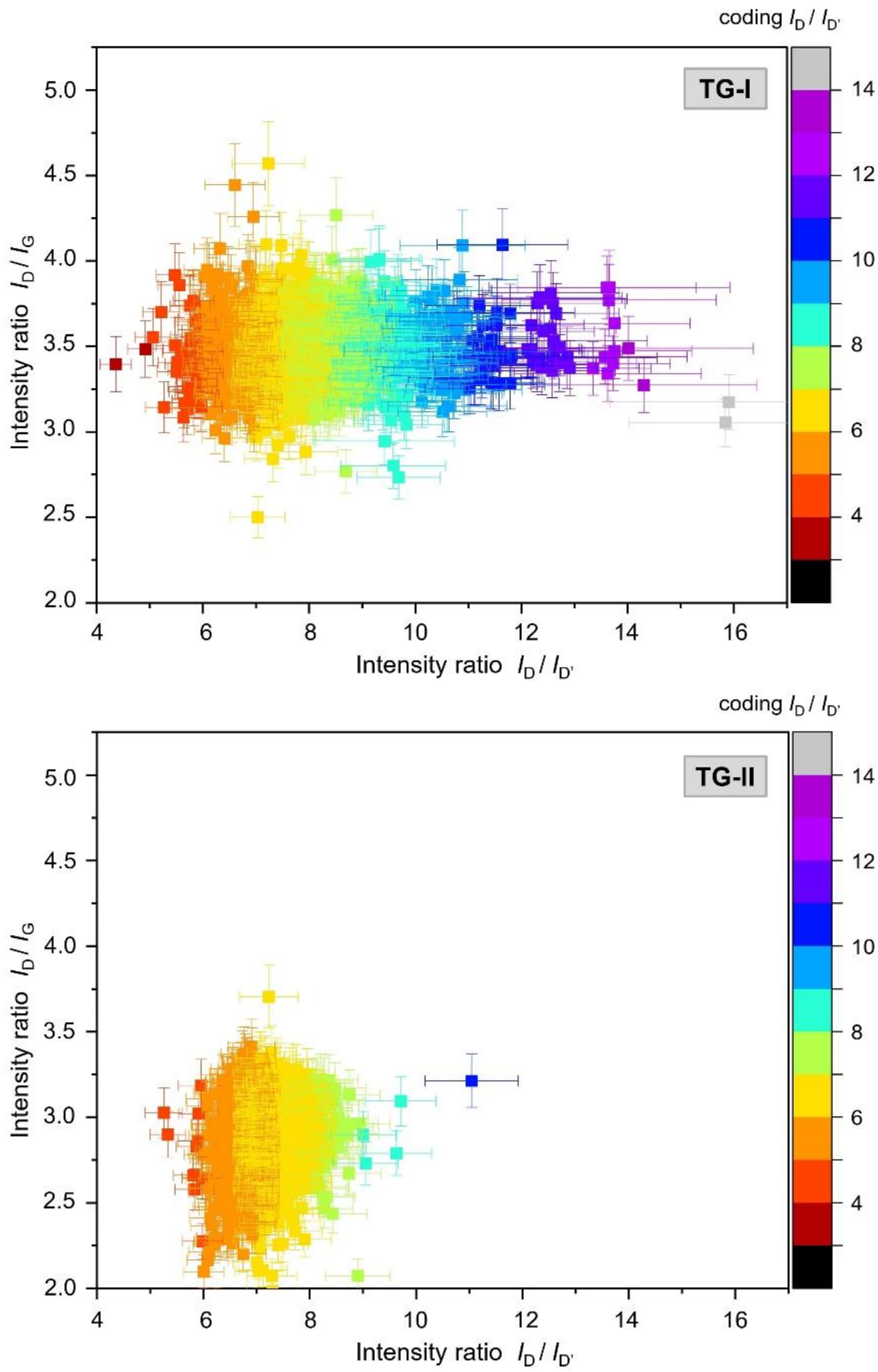

**Figure S4**: Intensity ratio plots of $I_D / I_G$ versus $I_D / I_{D'}$ of the 40×40 data points from the Raman maps in Figures S3(a) and S3(b), including error bars for each individual data point.

Note that in the figure the D/D' classification values are colour-coded according to integer-value ranges (analogous to the defect-type limiters).

For classification and interpretation, the values for each integer range are summed up, and the sum-values are shown as relative fractions of the total in Figure 6 of the main text. Said diagram serves as a visualisation aid in the discussion of the defect-type densities.

All (mean) evaluation parameters used for the presentation and interpretation of the experimental data – i.e., numerical values (including errors) for peak position, peak width and intensity ratios for the relevant Raman features G, D, D' and 2D – are summarised in Table S1.

**Table S1**: Mean fit results for all data sets. Given uncertainties are calculated from individual per spectra uncertainties, $\bar{u}_i$, and spread of the data points, $s$: $\sigma_{\text{tot}}=(\bar{u}_i^2 + s^2)^{1/2}$ . Note that, the individual uncertainties are much smaller than the spread of the data points within each data set, therefore, in first order, one can approximate $\sigma_{\text{tot}} \approx s$. The mixing parameter is defined in equation (4) of S2.

| | Peaks | Pristine | TG-I | TG-II | hTG-I | hTG-II |
|---|---|---|---|---|---|---|
| **Intensity ratio** | D/G | 0.12 ± 0.01 | 3.51 ± 0.26 | 2.90 ± 0.28 | 1.60 ± 0.24 | 1.53 ± 0.29 |
| | D'/G | 0.02 ± 0.01 | 0.49 ± 0.07 | 0.42 ± 0.05 | 0.38 ± 0.06 | 0.37 ± 0.11 |
| | D/D' | 6.22 ± 1.50 | 7.18 ± 1.00 | 6.94 ± 0.56 | 4.22 ± 0.50 | 4.26 ± 0.32 |
| | 2D / G | 1.75 ± 0.10 | 0.70 ± 0.09 | 1.28 ± 0.27 | 1.36 ± 1.08 | 1.15 ± 0.16 |
| | D / 2D | 0.07 ± 0.01 | 5.07 ± 0.57 | 2.32 ± 0.37 | 1.22 ± 0.31 | 1.41 ± 0.71 |
| **FWHM$_{(L)}$ in cm$^{-1}$** | D ($w_D$) | 54.2 ± 5.5 | 28.8 ± 1.3 | 27.4 ± 1.2 | 29.3 ± 3.5 | 31.8 ± 7.1 |
| | G ($w_G$) | 16.3 ± 0.9 | 39.7 ± 6.2 | 33.8 ± 3.4 | 21.3 ± 4.0 | 26.5 ± 6.0 |
| | D' ($w_{D'}$) | 21.8 ± 17.5 | 21.1 ± 6.5 | 22.4 ± 2.7 | 32.8 ± 8.7 | 23.6 ± 3.6 |
| | 2D ($w_{2D}$) | 32.9 ± 0.7 | 41.6 ± 6.6 | 44.8 ± 6.7 | 37.4 ± 4.3 | 39.0 ± 6.0 |
| **Position in cm$^{-1}$** | D | 1344.7 ± 1.2 | 1344.5 ± 3.3 | 1342.9 ± 0.8 | 1329.3 ± 1.3 | 1339.6 ± 4.7 |
| | G | 1584.5 ± 0.9 | 1581.9 ± 3.9 | 1581.0 ± 1.3 | 1572.6 ± 1.5 | 1583.4 ± 5.6 |
| | D' | 1621.1 ± 5.3 | 1622.7 ± 4.0 | 1621.8 ± 1.3 | 1601.3 ± 1.5 | 1618.7 ± 5.7 |
| | 2D | 2677.0 ± 1.8 | 2674.0 ± 6.7 | 2672.7 ± 1.6 | 2652.2 ± 2.6 | 2673.8 ± 20.3 |
| **Mixing parameter η** | D | 0.98 ± 0.01 | 0.93 ± 0.02 | 0.92 ± 0.01 | 0.92 ± 0.09 | 0.93 ± 0.03 |
| | G | 0.83 ± 0.01 | 0.95 ± 0.03 | 0.94 ± 0.01 | 0.87 ± 0.09 | 0.91 ± 0.04 |
| | D' | 0.83 ± 0.11 | 0.86 ± 0.07 | 0.89 ± 0.02 | 0.92 ± 0.10 | 0.89 ± 0.04 |
| | 2D | 0.94 ± 0.01 | 0.96 ± 0.03 | 0.97 ± 0.00 | 0.94 ± 0.09 | 0.95 ± 0.04 |

### *S3.2 Concept for the determination of the concentrations of defect types*

By and large, defect types (x) may be separated into two broad classes, namely adsorption-type (x≡a) and vacancy-type (x≡v) defects. According to established, descriptive models for defects in graphene they manifest themselves in changes in the amplitude and shape of Raman spectral features. For stage-1 defect-density samples, i.e., samples with small to moderate defect densities, one can take away two key messages (based on said models):

(i) the intensity ratio $I_D/I_G$ is proportional to the defect density, $L_D$(x), in units of (nm); and

(ii) the intensity ratio $I_D/I_{D'}$ yields distinct associations: $I_D/I_{D'} \geq 13$ is indicative for adsorption defects; $I_D/I_{D'} \leq 7$ for vacancy defects; and $I_D/I_{D'}$ in the range 7⋯13 for mixed defects.

Note however that, mixing cannot necessarily be assumed to be linear.

The established model for vacancy-type defects is that by Lucchese and co-workers,[32] which was generalized for different laser excitation energies, $E_L$ (in eV), by Cançado and co-workers.[33] The latter describes the relationship between $I_D/I_G$ and $L_D$(v) as

$$\frac{I_D}{I_G}(L_D) = C_A \frac{r_A^2 - r_S^2}{r_A^2 - 2r_S^2} \cdot \left[ e^{-\pi \frac{r_S^2}{L_D^2}} - e^{-\pi \frac{r_A^2 - r_S^2}{L_D^2}} \right] + C_S \left[ 1 - e^{-\pi \frac{r_S^2}{L_D^2}} \right] \tag{5}$$

with $C_A = 160 \cdot E_L^{-4} \cdot \text{eV}^{-4}$, $C_S = 0.87 \pm 0.05$, $r_A = (3.00 \pm 0.03)\text{nm}$ , and $r_S = (1.00 \pm 0.04)\text{nm}$ .

Recently Fournier and co-workers published a model specifically tailored for sp$^3$-type defects.[36] There the relationship between $I_D/I_G$ and $L_D$(a) is defined by

$$\frac{I_D}{I_G}(L_D) = \frac{C_S f_S(L_D) + C_A f_A(L_D)}{1 - f_{sp^3}(L_D)} \; , \tag{6}$$

where

$$f_A(L_D) = e^{-\pi \frac{r_S^2}{L_D^2}} - e^{-\pi \frac{r_A^2}{L_D^2}} \, , \; f_S(L_D) = e^{-\pi \frac{r_{sp^3}^2}{L_D^2}} - e^{-\pi \frac{r_S^2}{L_D^2}} \text{ and } f_{sp^3}(L_D) = 1 - e^{-\pi \frac{r_{sp^3}^2}{L_D^2}} \; , \tag{7}$$

and $C_A = 20.1$, $C_S = 0.86$ , $r_A = 0.242\text{ nm}$ , $r_S = 0.183\text{ nm}$ and $r_{sp^3} = 0.0913\text{ nm}$ .

When evaluating and comparing our data – using these two models – one finds that, the Raman response to the two defect types is significantly more sensitive for vacancy-type defects than for adsorption-type defects. For example, a value of $I_D / I_G = 0.1$ would correspond to

$L_D$(Lucchese / Cançado) = 37.1 nm , or

$L_D$(Fournier) = 4.1 nm .

This means that, when quantitatively comparing data sets, for compositional changes, both variations in $I_D / I_G$ and $I_D / I_{D'}$ need to be tracked in parallel.

### *S3.3 Example calculation*

Simplified versions of the complex equations (1) and (2) can be found in the literature, which allow for an analytical inversion; however, these are in general only applicable in the low-defect density regime. Since this is not the case for the samples studied in this work, we use a different approach.

In order to accurately evaluate the defect density, $L_D$, from the intensity ratio, $I_D / I_G$, we calculate the correlation curves according to the two models, and then perform numerical interpolation. This principle is visualised for the TG-I data in Figure S5.

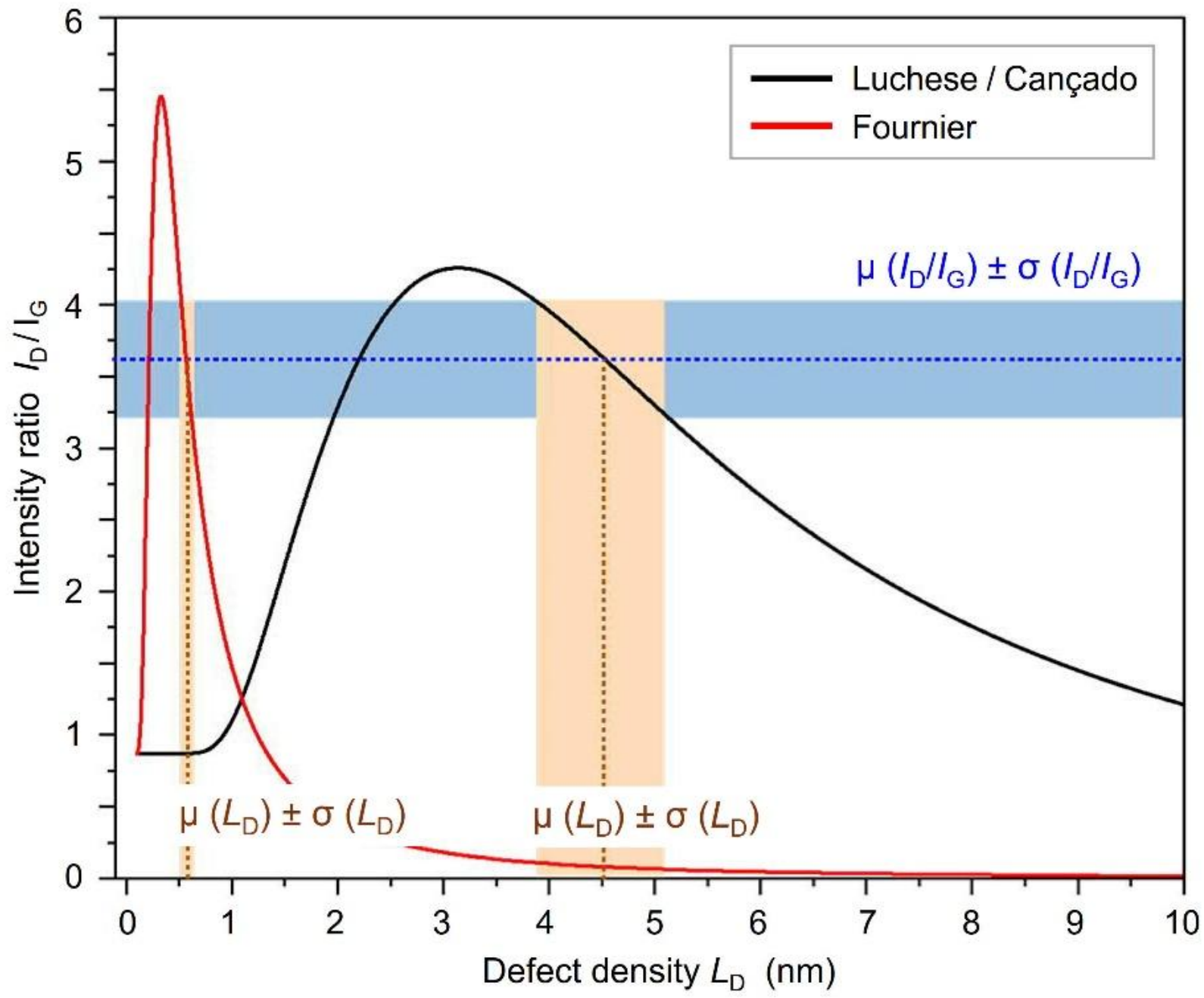

**Figure S5**: Calculated numerical $I_D/I_G$ vs $L_D$ curves, according to the models of Lucchese/Cançado and Fournier. The blue dotted line and the blue band correspond to the mean $I_D/I_G$ with 1σ standard uncertainty of the TG-I data set. As it is shown in the main document, the TG-I data fall to the low-defect regime (Stage 1); therefore, the crossing as smaller $L_D$-values is correct here. The orange lines and bands visualise the $L_D$-values and their uncertainties; they are obtained by numerically interpolating the intersection points with the blue line, and their uncertainty, for each of the two model curves.